# Compensation effect and magnetostriction in $CoCr_{2-x}Fe_xO_4$

**Hong-guo Zhang, Weng-hong Wang**[*]**, En-ke Liu, Xiao-dan Tang, Gui-jiang Li, Hong-wei Zhang, Guang-heng Wu**

State Key Laboratory of Magnetism, Institute of Physics and Beijing National Laboratory for Condensed Matter Physics, Chinese Academy of Sciences, Beijing 100190, China

**Keywords**:   $CoCr_{2-x}Fe_xO_4$, compensation, magnetostriction, multiferroic

**Abstract**
The magnetic compensation and magnetostriction properties in Fe doped $CoCr_2O_4$ samples have been investigated. Structural and magnetic measurements imply that the doped $Fe^{3+}$ ions initially occupy the B1(Cr) sites when x<0.1, and then mainly take the A(Co) sites. This behaviour results in a role conversion of magnetic contributors and a composition compensation between two competitively magnetic sublattices at x=0.1. Temperature dependence compensation has also been found in the samples with x=0.1~0.22, with the compensation temperature in the range of 40~104 K. The $Fe^{3+}$ doping also modulates the exchange interaction of the system and prevents the formation of long range conical order of spins. The magnetoelectric transition temperature at 23K in $CoCr_2O_4$ is shifted to lower temperature by increasing the dopants. The magnetostriction effect in this system has been observed for the first time. The strain has a maximum value of about 280ppm at x=0.4. The magnetostriction is in consistent with the behaviour of the two magnetic compensations.

## 1 Introduction

Multiferroics are the materials that possess both ferroelectricity and ferromagnetism in the same phase, which have attracted great attention due to their physical interest and several potential technological applications[1-3]. Single phase multiferroics are very rare and could be classified into several groups based on their microscopic origin[4,5]. Among them, a number of frustrated magnets are found to exhibit magnetoelectric coupling with the emerging of ferroelectricity induced by their spiral spin structures[6-10]. Spinel CoCr2O4 which shows a unique conical-spiral ferrimagnetic spin order[11] is the first example of a multiferroic with both a spontaneous magnetization and ferroelectric polarization of spin origin[12].

$CoCr_2O_4$ is a normal spinel that $Co^{2+}$ ions occupy the tetrahedral (A) sites and magnetic $Cr^{3+}$ ions occupy the octahedral (B) sites[11]. The system experiences a paramagnetic to ferrimagnetic

---

[*] Corresponding author: e-mail wenhong.wang@aphy.iphy.ac.cn Phone: +86 82 649 247

transition at 97 K, then starts a spiral component short-range order which transforms into long-range order at 26K ($T_s$)[12,13]. Generally, It is believed that the dominant origin of the ferroelectric polarization in $CoCr_2O_4$ is from the conical order on $Cr^{3+}$ sublattice[12]. However, there are evidences that the dipole moments of $Co^{2+}$ sublattice on A-sites should not be neglected at low temperature[14,15]. The conical orders on these sublattices are very sensitive to the nearest neighbor and isotropic antiferromagnetic A-B and B-B exchange interactions. Vary the cations on the A/B sites would change the cone angle of the conical order and thus influence the ferroelectric polarization relied on them[13,16]. Therefore, it is conceivable that the doping on A/B sites may interrupt the delicately balanced of the exchange interaction of the cations and modulates the magnetic and magnetoelectric properties.

Among the works on the cation substitution of $CoCr_2O_4$ system[17-20], there are few reports concerning the magnetic properties when the $Cr^{3+}$ ions is substituted by Fe[18]. Considering that $Fe^{3+}$ ions have stronger spin moment than $Cr^{3+}$, introducing of $Fe^{3+}$ on B-sites may cause the enhancement of magnetization and affect the multiferroic properties in $CoCr_2O_4$. In the present work, a series of Fe doped $CoCr_2O_4$ samples have been synthesized and the systematic examination about structural, magnetic and magnetostrictive properties have been performed. It is found that the distribution of doped $Fe^{3+}$ ions results in the apparent composition and temperature dependence compensation effects. The observation of magnetostriction in the series has also been reported in the present work.

**2 Experimental details**

A series of polycrystalline Fe-doped cobalt chromite samples with compositions of $CoCr_{2-x}Fe_xO_4$ (x=0-1) were prepared in air by the solid-state reaction method using $Fe_2O_3$ (99.5%), $Cr_2O_3$ (99%), and $Co_3O_4$ (99.7%) powders as precursors. The initial mixtures were well ground, pelletized, sintered twice at 1200℃ for 24h, and were subsequently furnace cooled to room temperature.

The samples were examined by powder XRD (X-ray diffraction) measurements using Cu$K$α radiation (Rigaku Co., RINT-2400) with an angle (2θ) step of 0.02 between 2θ=10° and 90°. The EXAFS measurements were performed at the 1W1B beamline at Beijing Synchroton Radiation Facility (BSRF). The dc magnetization measurements were carried out on a superconducting quantum interference device (SQUID, Quantum Design MPMS XL-7). Magnetization temperature (MT) curves were measured in field cooled (FC) and zero-field cooled (ZFC) modes with various applied field ($H$) and temperature ranges. The specific heat against temperature was obtained by using the heat capacity unit on Physical Property Measurement System (PPMS). Magnetostriction as a function of applied field ($\lambda/H$) and temperature ($\lambda/T$) were measured using the strain gauge method on PPMS.

**3 Result and discussions**

All the samples have cubic spinel structure confirmed by using x-ray powder diffraction, as shown in figure 1. The lattice parameter against x is illustrated in figure 1(a). It monotonously increases with increase of $x$, indicating the continuous formation of solid solutions. This is due to that both end compounds ($CoCr_2O_4$ and $CoFe_2O_4$) are with cubic spinel structure in *Fd-3m/Fd3m* space groups[21,22]. The interesting thing is that, as shown in figure 1(b), the variation of lattice parameter shows two sections with different slopes at x = 0.8~1.0. The overall increase of lattice parameter is understandable because the ionic radii of $Fe^{3+}$ in octahedron (0.785Å) is larger than that of $Cr^{3+}$(0.755Å). However, the doped Fe ions usually prefer to take the A-sites instead of B-sites, as in $CoFe_2O_4$, and form a inverse spinel structure[23]. Therefore, one could expect that the doped Fe ions mainly distribute on A-sites as the increase of x and the replaced Co ions will occupy the B-sites. Considering that the Co ion has a radii of 0.885Å in octahedron, even larger than it of $Fe^{3+}$, thus the linearly increase of the lattice parameter above x=1.0 is relatively rapid. As for the slow increase between x=0.0~0.8, it may suggest that the doped Fe ions firstly occupy the B-sites and causes a gently enhancement of the lattice parameter. Further confirmation of the occupation on A-sites of Fe ions will be discussed later.

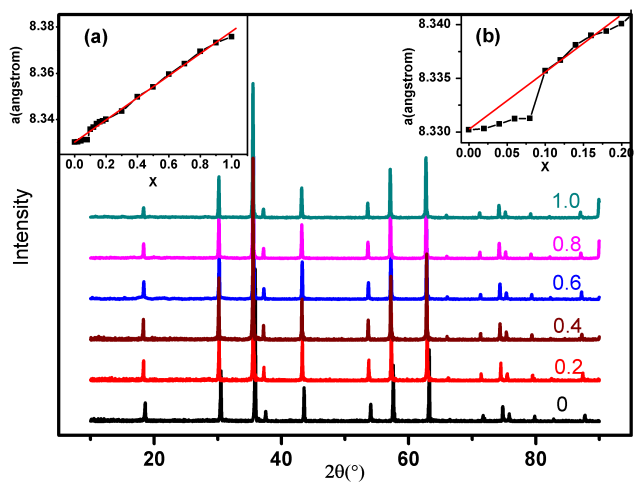

**Figure 1** The main plate shows the XRD patterns of $CoCr_{2-x}Fe_xO_4$ (0≤x≤1); the inset (a) shows the lattice parameters *a* as a function of *x* and (b) presents the enlarged view of the range from 0 to 0.2.

EXAFS experiments have been carried out in this work to identify the distribution of the doped Fe ions. As shown in figure 2(a), the spectrums of doped samples with x = 0.5 and 1.0 are very similar to that of x = 2 ($CoFe_2O_4$). According to the previous reports[24-26], only if the Fe ions appear on tetrahedral A-sites without inversion center of symmetry, there shows the pre-edge peak like in Fig.2. As shown in figure 2(b), the intensity of pre-edge peak grows with increase of Fe ions, which reflects the amount of $Fe^{3+}$ occupation on tetrahedral A sites in this doped system. These results further prove that the doped $Fe^{3+}$ ion dominantly distributes on the A site in most of the doping range, except x = 0~0.1, as indicated above by the XRD examination.

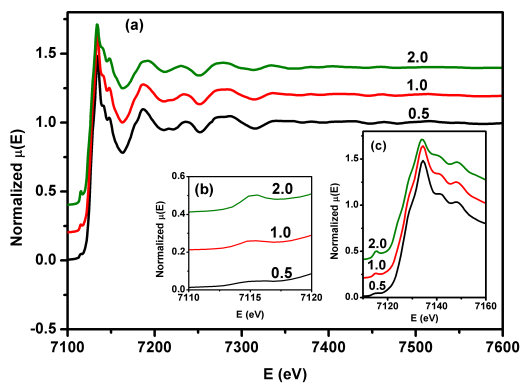

**Figure 2** (a) Extend x-ray absorption fine structures (EXAFS) of Fe K-edge at room temperature for $CoCr_{2-x}Fe_xO_4$ (0.5, 1.0 and 2.0); (b) the pre-edge area of the EXAFS and (c) the enlarged view of the near edge

The EXAFS spectrums also give us an important clue about the valency of Fe dopants. From figure 2 (b), one can see that the absorption edges of the samples gradually move to low energy along increasing x. The oxidation number of Fe dopants can be estimated using the Fe K- edge threshold energy, i.e., the energies corresponding to the peaks of 1st derivative of μ(E)[27,28]. Since Fe ions in $CoFe_2O_4$ have effective valences less than +3, the shift of edge for the doped samples to lower energy side implies the relatively high oxidation.

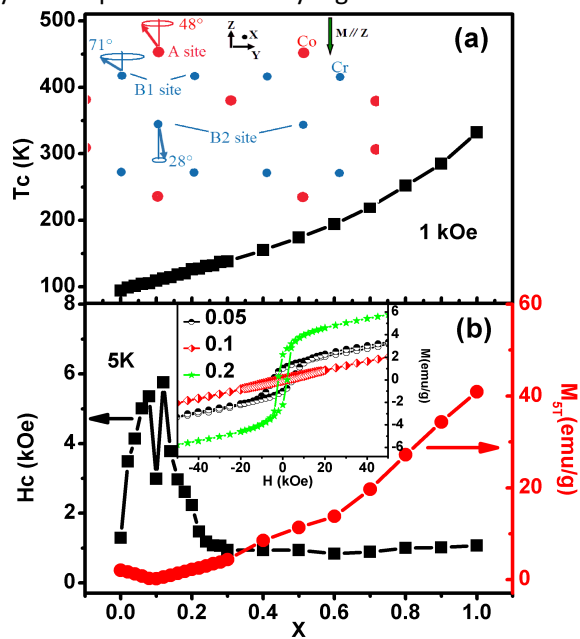

**Figure 3** Composition dependences of (a) Curie temperature (Tc), (b) coercive force (Hc) and magnetization under 5T ($M_{5T}$). The inset of (a) is the illustration of the cation distribution in $CoCr_2O_4$, view from the x axis. The inset of (b) is the field dependences of magnetization (MH) for several samples

The temperature and field dependences of magnetization of the series have been measured. Figure 3(a) shows that the Curie temperature ($T_C$) enhances monotonously with increase of x, which implies the strengthening of exchange interaction despite the expansion of lattice. The magnetization dependence measured under 5T ($M_{5T}$) shows a minimum at x=0.1, as shown in figure 3(b), indicating a composition compensation effect. Interestingly, approaching to the compensation point from the both sides, the coercivity ($H_C$) shows an apparent increase, but sharply drops to a minimum value of 2986 Oe at x = 0.1. Considering the particular distribution of the dopants, these phenomena will be discussed below in the view of the competition between doped magnetic sublattices.

The inset of figure 3(a) shows the magnetic structure of $CoCr_2O_4$ projected on the YZ plane[12,13]. Within A(Co) and B(Cr) sublattices, the moments on A and B1/B2 sites are conically ordered and their cone axes are antiparallelly aligned to each other. The ferrimagnetic components of the conical spins along the cone axes induce a macroscopic moment in the direction of z [001]. Considering the spin-only ionic magnetic moments of $Fe^{3+}(5\mu_B)$, $Cr^{3+}(3\mu_B)$ and $Co^{2+}(3\mu_B)$, only if B1 and A sites, rather than B2 sites, are taken by the doped Fe, the magnetic compensation would occur. Remind the conclusion derived from the XRD and EXAFS results, while x<0.1, the dominant occupation of Fe ions should be on B1 sites and leads to the composition compensation. That is to say the main contribution to magnetization in the ground state of the series comes from B sites when x<0.1 and turns to the A sites while x > 0.1. Further investigation on the temperature dependence of magnetization will reveal that the main contributor switch between two sublattices during warming/cooling process.

As to the variation of $H_C$, the sharply increase of $H_C$ while x<0.1 may imply that the introduction of $Fe^{3+}$ on B1 sites enhances the anisotropy of the system. However, the minimum of Hc at 0.1 may be attributed to the abnormal shape of MH loop due to the near zero magnetization at the compensation point. This had also been observed in other systems[29,30]. When x>0.1, the doped A sublattice becomes domination, inducing the decrease of the $H_C$ which shows a constant value after x=0.3.

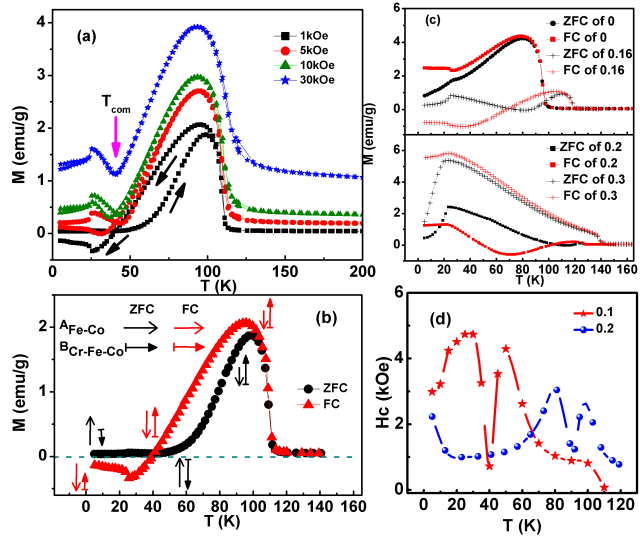

**Figure 4** (a) the ZFC and FC curves of x=0.1 for 1, 5, 10, 30 kOe, respectively; (b) illustration of the compensation and negative magnetization in x=0.1; (c) ZFC-FC curves of several samples under 1kOe; (d) temperature dependence of Hc for x=0.2.

Figure 4(a) shows the ZFC-FC curves of the sample with x=0.1 measured under different fields. One may found that a negative magnetization appears on the FC curve measured under 1kOe at about 40K. Because this field is much lower than $H_C$ and the energy barrier of anisotropy is too high to be overcome by the external field, there is no response of the domain walls during the temperature change. Therefore, the negative value of magnetization implies a role conversion for magnetic contributor occurs through 40K (labelled as $T_{com}$), showing a temperature compensation behaviour.

In the case of applying 30kOe field, the curve shows that the minimum has a non-zero value at $T_{com}$. This strongly suggests the magnetic structure has been changed and the cone axes of the two sublattices are non- collinearized by a high field. While applying field of 5 and 10kOe, minimums on ZFC and FC curves show a thermal hysteresis. Counting in the magnitude of $H_C$ shown in figure 3(b), it is reasonable to believe that the domains can not been switched completely by a field within this range. The compensation behaviour may still be affected by the pinning of the domain walls.

The mechanism for magnetic contributor changing is illustrated in figure 4(b) related to the moments in two sublattices. In this Figure, two magnetic sublattices are defined as $A_{Fe-Co}$ (Fe and Co atoms on A sites) and $B_{Cr-Fe-Co}$ (Cr, Fe and Co on B sites). Notice that the ZFC measurement shows a flat curve through $T_{com}$. This is due to the freeze of domains in a high anisotropy system in which the alignment of the moments by external field becomes very difficult and the role conversion for magnetic contributor is not apparent.

Figure 4(c) shows the *M-T* curves for the samples with x = 0 ~ 0.3. The figure of $CoCr_2O_4$ shows no compensation effect, indicating the domination of $B_{Cr-Fe-Co}$ sublattice in the whole temperature range. The negative magnetization has also been observed in the samples with x = 0.16 and 0.20, Their $T_{com}$ increases with the increase of x, which implies the continuously strengthening of the $A_{Fe-Co}$ sublattice. With cooling down the temperature further below $T_{com}$, FC curve turns to positive again for the sample of x=0.2. It comes from the continuously decreases of $H_C$ in this temperature range, as shown in figure 4(d), where the anisotropy is comparable with the external field. With more Fe content of x=0.3, the curve shows a monotonously increase before the magnetoelectric transition, which means that the domination contributor becomes the $A_{Fe-Co}$ sublattice. The magnetic structures in these samples are likely to be more colinear as in $CoFe_2O_4$, duing to the effect of Fe doping on the exchange interaction between the A and B sites. Figure 4(d) shows the temperature dependence of $H_C$ of the samples with x = 0.10 and 0.20. Similar to the case of the composition compensation, the minimum of $H_C$ has also been observed in the temperature compensation situation at $T_{com}$.

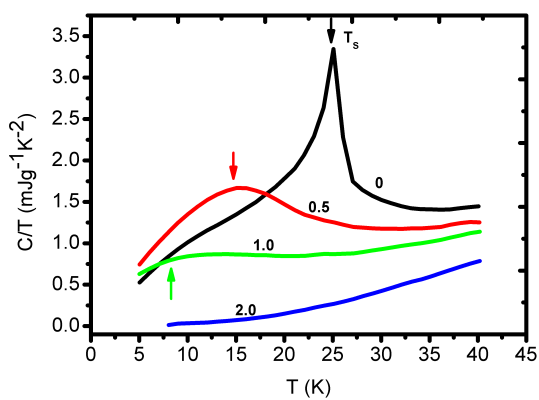

**Figure 5** Specific heat over temperature curves of the series from 5K to 40K.

The specific heat over temperature (*C/T*) for several samples was carried out to identify the multiferroic transition temperatures of the samples, showing in figure 5. Increasing the Fe content, the C/T peak at $T_s$ is moving to the lower temperature, indicating that the chemical doping on B-site results in suppression of the conical spin structure. This may due to two scenarios: the $Fe^{3+}$ ions on B site tend to break the long range order of spiral structure and the A-site $Fe^{3+}$ ions enhance the strength of the A-B interaction and thus weakens the frustration on B sites. Besides, the intensity of the C/T peak decreases and lasts in larger temperature range with increased doping. This indicates the diffusion of the magnetic transition and unsaturation of the coherent length of conical order. The transition is fully suppressed above x=1.0 whereafter the magnetic structure turns to more collinear like in $CoFe_2O_4$.

Figure 6 shows the composition dependence of magnetostriction for the series under 5K. It has been found that our $CoCr_{2-x}Fe_xO_4$ system exhibits a quite large magnetostrictive value of up to about 280 ppm. This property had never been investigated so far. As shown in figure 6, the magnetostriction shows a minimum at x = 0.10 and a maximum at about 0.40, being consistent to the compensation behaviour. This observation indicates that there is a magnetostriction existed natively in $CoCr_2O_4$, even though, as can be see of the inset of figure 6, this strain originated from the conical structure does not have the same style with the magnetization. The unsaturation of the strain against field could be attributed to the noncollinear magnetic structure in the system, which is also difficult to fully align along the external magnetic field. In the doping range of x < 0.1, the native magnetostriction is reduced by doped $Fe^{3+}$ through suppressing the conical order, as the situation observed in the specific heat. This induces the continuously weak of the coupling between spiral order and lattice and thus the decrease of the strain. The consistent minimums of strain and lattice parameter as well as magnetization at the same component (x=0.1) suggests that they all related to the compensation effect at this point. While x > 0.1, the $Co^{2+}$ ions on B sites introduce another apparent magnetostriction effect through their incompletely quenched orbital angular momentum[31], resulting in a growth of magnetostriction below x = 0.40. This can also be reflected from the inset of figure 6, where the low field slope of the strain enhances sharply for the curves of samples with x>0.1. After x = 0.40, the strain begins to de- crease, which implies a serious interruption of this ordered magnetic structure.

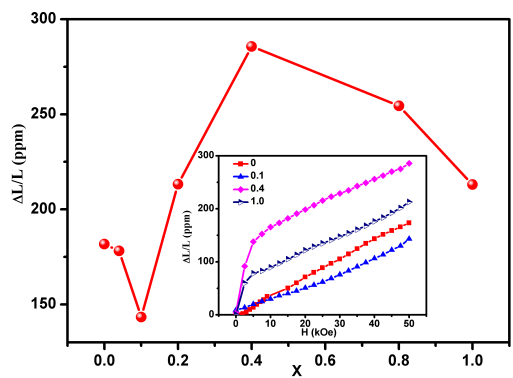

**Figure 6** The composition dependence of magnetostriction under 5T. The inset shows the curves of strain against external field for different samples.

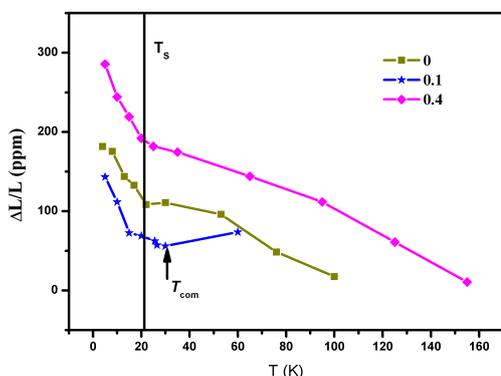

**Figure 7** The temperature dependences of magnetostriction under 5 kOe of the series.

The temperature dependences of strain under 5 kOe in the series are shown in figure 7. One can see that the strain undergoes the similar changes at the transition and compensation temperatures as in the MT curves. Under $T_S$, there obviously is a strong coupling between the long order of spiral structure and the crystal lattice, leading to a rapid increase of magnetostriction. While above $T_S$, due to lacking of long range order, the strain continuously reduces with the increasing of temperature. Besides, the magnetostriction also shows a minimum at $T_{com}$, implying that the strain below and above this temperature is derived from two different magnetic sublattices dominant at different temperature ranges.

## 3 Conclusion

In conclusion, the magnetic compensation and magnetostriction of CoCr$_{2-x}$Fe$_x$O$_4$ (x=0-1) series have been studied. We find out that the doped Fe$^{3+}$ ions start to occupy B1 sites initially when x<0.1, and then prefer to take A sites instead. This results in a role conversion of magnetic contributors and a composition compensation between two competitively magnetic sublattices at x=0.1. In the samples with x=0.1~0.22, a temperature dependence compensation can be found in the temperature range of 40~104K. The long range conical spin order is suppressed by doping, with $T_S$ shifts to low temperature and the C/T peak becomes weak. For the first time, the magnetostriction effect with a strain of 280 ppm has been observed in the sample with x=0.4 at 5K. It has also been found that the magnetostrictive property is in consistent with the composition and temperature compensation behaviour.

**Acknowledgements** This work was supported by the National Basic Research Program (2009CB929501), specific funding of Discipline & Graduate Education Project of Beijing Municipal Commission of Education, and the National Natural Science Foundation of China in Grant No. 11174352.